# Transport Evidence of Magnetic Polarization in the Altermagnetic Candidate MnTe

Younes Ghorbani,[1] Nayana Devaraj,[2] Joshua Maile,[1] Samuel Poage,[1] Qihua Zhang,[3] Maria Hilse,[3] Stephanie Law,[3] Salva Salmani-Rezaie,[1] Awadhesh Narayan,[2] and Kaveh Ahadi [1,4,a)]

1) Department of Materials Science and Engineering, Ohio State University, Columbus, OH 43210 USA

*2) Solid State and Structural Chemistry Unit, Indian Institute of Science, Bangalore 560012, India*

3)Two-Dimensional Crystal Consortium Materials Innovation Platform, The Pennsylvania State University, University Park, Pennsylvania 16802 USA

4) Department of Electrical and Computer Engineering, Ohio State University, Columbus, OH 43210 USA

a) ahadi.4@osu.edu

**Abstract**

The ability to precisely control magnetic properties is central to the development of future spin-based electronics. In this work, we report the successful growth of epitaxial α-MnTe thin films on InP(111) substrates using molecular beam epitaxy. Magneto-transport measurements at low temperatures reveal a distinct, hysteretic butterfly longitudinal magnetoresistance alongside a nonlinear transverse magneto-resistance response, suggesting the presence of a finite net magnetic polarization in the films. To understand the origin of this behavior, density functional theory (DFT) calculations were performed. While pristine bulk MnTe is a compensated antiferromagnet, our computational results suggest multiple pathways through which a finite magnetization can emerge in thin-film geometries, including interface-induced symmetry breaking and point defects. These findings demonstrate an epitaxial route for engineering magnetic responses in thin films.

Precise control over magnetic properties is essential for advancing next-generation spin-based devices, where performance depends on magnetic anisotropy, ordering temperature, and spin dynamics[1–3]. Among the various approaches, epitaxial growth offers a powerful route to engineer magnetism by tailoring crystal orientation, lattice strain, and defect density. Such epitaxial control enables systematic tuning of exchange interactions and magnetic anisotropy, providing access to magnetic states that are often unattainable in bulk materials[4]. Understanding and exploiting epitaxial structure–property relationships is a key strategy for realizing magnetic materials with controllable functionalities.

α-MnTe crystallizes in the NiAs-type hexagonal structure ($P6_3/mmc$ space group), comprised of alternating Mn and Te planes with *AB* stacking along the crystallographic *c*-axis [Figure 1(a)][5]. In this structure, Mn and Te atoms occupy the 2a ($\bar{3}$m symmetry) and 2c ($\bar{6}$m2 symmetry) Wyckoff positions, respectively. The magnetic ground state of MnTe energetically favors an *A*-type antiferromagnetic configuration, where the spins of Mn atoms are aligned parallel within each basal plane, while the magnetic sublattices along the *c*-axis are aligned antiparallel to each other[6–8].

Accordingly, net magnetization is expected to vanish, as each unit cell contains two Mn atoms with identical magnetic moments but opposite directions [Figures 1(a) and (b)]. The two antiferromagnetically coupled Mn sublattices are related by a non-symmorphic screw-rotation symmetry involving a sixfold rotation about the crystallographic *c*-axis accompanied by a fractional translation. Figure 1(c) shows that, despite the absence of net magnetization, the spin degeneracy of electronic bands is lifted along the L-Γ-L′ [Figure 1(d)]. Figures 1(e) and (f) exhibit the six-fold rotation connection of magnetization density of Mn atoms in α-MnTe. This symmetry relation precludes the restoration of effective time-reversal symmetry in momentum space and consequently allows the spin degeneracy of electronic bands to be lifted. The observed momentum-dependent spin splitting in Figure 1(c), along the L′-Γ-L, originates from the magnetic-crystallographic symmetry intrinsic to α-MnTe and occurs without the presence of any spin canting or extrinsic defect-induced magnetic moments. This symmetry-driven electronic response is hallmark of the altermagnetism in MnTe and clearly distinguishes it from conventional collinear antiferromagnets[9–12].

A tunable magnetic response where we could control magnetic polarization and time-reversal symmetry breaking by synthesis parameters is of interest. Here, we report on epitaxial synthesis of MnTe heterostructures on InP (111) single crystal substrates using molecular beam epitaxy (MBE). The films show a near room temperature magnetic transition and electronic transport signatures of time-reversal symmetry breaking. Density functional theory investigates the potential intrinsic and extrinsic causes of the observed magnetic polarization.

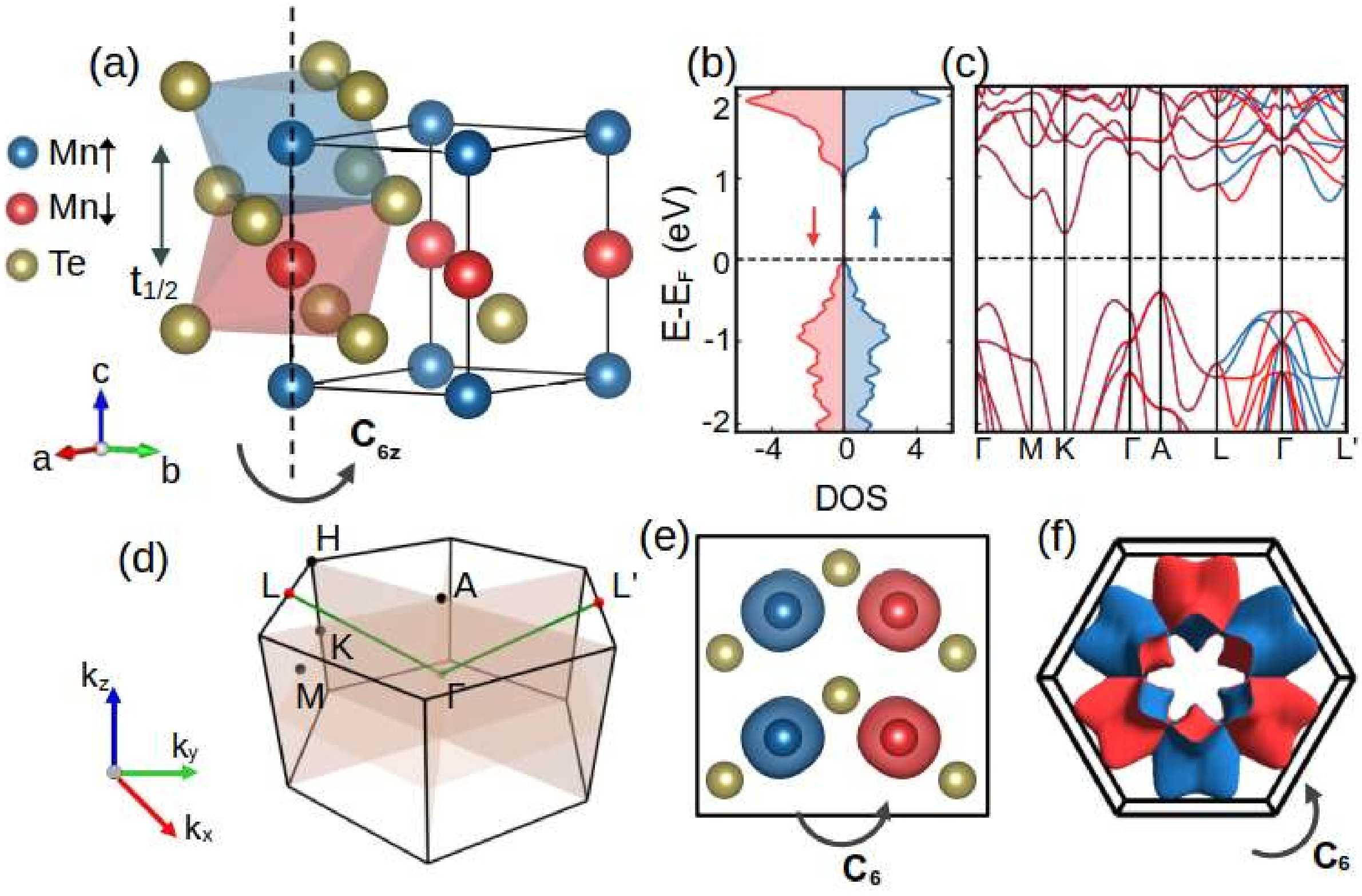

Figure 1. Magnetic structure of α-MnTe. (a) Crystal structure of α-MnTe, showing the two opposite-spin sublattices in blue and red. Opposite spin sublattices are connected by a sixfold screw-axis rotation. (b) Spin-polarized density of states with total density of states for up and down spin channels. (c) Spin-polarized electronic band structure along the high-symmetry path *Γ-M-K-Γ-A-L- Γ-L'*. Here, blue and red lines indicate the spin-up and spin-down bands, respectively. Spin-splitting is clear along the *L-Γ-L'*, while the bands for the two spin channels are degenerate along the Γ-M-K-Γ-*A*-L. (d) Bulk hexagonal Brillouin zone of α-MnTe, where the high-symmetry points denoted. The momentum-path where spin-splitting is seen is marked by green lines. (e) Opposite-spin sublattices connected by a sixfold rotation along the *c*-axis with magnetization density distribution. (f) A sixfold rotation connects the constant-energy surfaces for up and down spins, plotted at $E\text{-}E_F$=1 eV.

MnTe thin films were grown on semi-insulating Fe-doped InP (111)A substrates using a chalcogenide MBE system (DCA R450). Prior to the growth, the substrates were cleaned in isopropanol ultrasonic bath and thermally outgassed at 200 °C for 2 h in the load-lock chamber. The substrates were then annealed at 650 °C for 10 min under a Te flux of $2.3\times10^{14}$ $cm^{-2}$ $s^{-1}$ before being cooled to the MnTe growth temperature of 380 °C. MnTe was deposited using a Te/Mn flux ratio of 3 for 10 min, corresponding to a thickness of ~15 nm, corroborated by X-ray reflectivity (XRR) [Figure S1(c)][13]. In-situ Reflection high-energy electron diffraction (RHEED) exhibits streaky patterns suggesting a two-dimensional growth with smooth film surface [Figure S2]. After growth, the sample was held at the growth temperature under Te flux for 2 min before cooling to room temperature and capped with Si at room temperature using an electron-beam evaporator (10 kV, 131 mA) until the RHEED pattern disappeared. Hall bars were fabricated using conventional photolithography. The magneto-transport measurements were carried out using a Physical Property Measurement System (DynaCool, Quantum Design). The crystal structure and epitaxial quality of the MnTe thin film were investigated using X-ray diffraction (XRD) and high-angle annular dark-field scanning transmission electron microscopy (HAADF-STEM).

For the computational part, electronic structure calculations were performed using the Quantum ESPRESSO package[14] within the framework of density functional theory (DFT). The calculations employed a plane-wave basis set together with projector augmented-wave (PAW) pseudopotentials[15]. Exchange-correlation effects were treated using the Perdew–Burke–Ernzerhof (PBE) functional within the generalized gradient approximation (GGA)[16]. A plane-wave kinetic energy cutoff of 60 Ry was used throughout the calculations. For the MnTe/InP heterostructures, a vacuum region exceeding 15 Å was introduced along the out-of-plane direction to eliminate interactions between periodically repeated images. Structural relaxations were carried out using a 5 × 5 × 1 Monkhorst–Pack *k*-point mesh, whereas a denser 20 × 20 × 1 mesh was employed for the electronic structure calculations. The self-consistent field convergence threshold was set to $10^{-9}$ Ry. To investigate the effects of vacancies, interstitials, and substitutional dopants, a 2 × 2 × 2 supercell of MnTe was constructed. For these calculations, a 5 × 5 × 3 *k*-point mesh was used during structural optimization, while a 10 × 10 × 6 mesh was adopted for the electronic structure calculations. To properly describe the localized Mn 3*d* electrons, the DFT+U approach[17] was employed. A Hubbard parameter of $U$ = 3 eV was applied to the Mn *d* orbitals, consistent with previous theoretical and experimental studies[6,18–21].

The θ-2θ XRD scan only exhibits *000ℓ* peaks [Figure S1(a)], confirming that the film is single-phase and oriented along the out-of-plane direction of the substrate. The rocking curve full width at half maximum (FWHM) of the MnTe film is 395 arcseconds [Figure S1(b)], suggesting high crystalline quality of the grown films[22–25].

Figure S3 exhibits the cross-sectional HAADF-STEM image. The imaging shows that the MnTe film is continuous over the imaged region, with no obvious extended defects observed in the film. Higher-magnification imaging of the interface indicates that the MnTe/InP interface is not atomically abrupt. This is consistent with the accompanying EDS maps and line profile, which shows a broadened chemical transition and overlap of the Mn/Te and In/P signals across the interface, suggesting a diffuse or partially intermixed interfacial layer. The interfacial steps in (111) heterostructures of other materials system also exhibit similar features[26,27]. Although the film appears structurally coherent in projection away from the interface, MnTe is air sensitive and the film degrades over time upon ambient exposure[22,28,29].

Figure 2 exhibits the temperature dependence of the longitudinal and transverse sheet resistance with temperature. We observe a peak in both longitudinal and transverse channels. The signal in transverse channel in the absence of applied magnetic field could suggest a broken time-reversal symmetry. Interestingly, the peaks in the longitudinal and transverse channels do not coincide. Here, a metallic like behavior is observed below 250 K in longitudinal resistance with temperature ( $dR_{xx}/dT > 0$), despite a reported ~1eV bandgap in MnTe [21,29,30]. This could be due to presence of point defects and deviation from ideal stoichiometry[29]. We attribute the resistance peak in longitudinal and transverse transport to a magnetic transition from the high temperature paramagnetic state. The Néel temperature of bulk MnTe is reported to be ~300 K[32]. The resistance peak does not necessarily mark the magnetic transition temperature, but scattering matrix of charge carriers could be affected by the magnetic transition. We also notice hysteretic behavior with temperature sweeping direction, suggesting a first order phase transition[33,34].

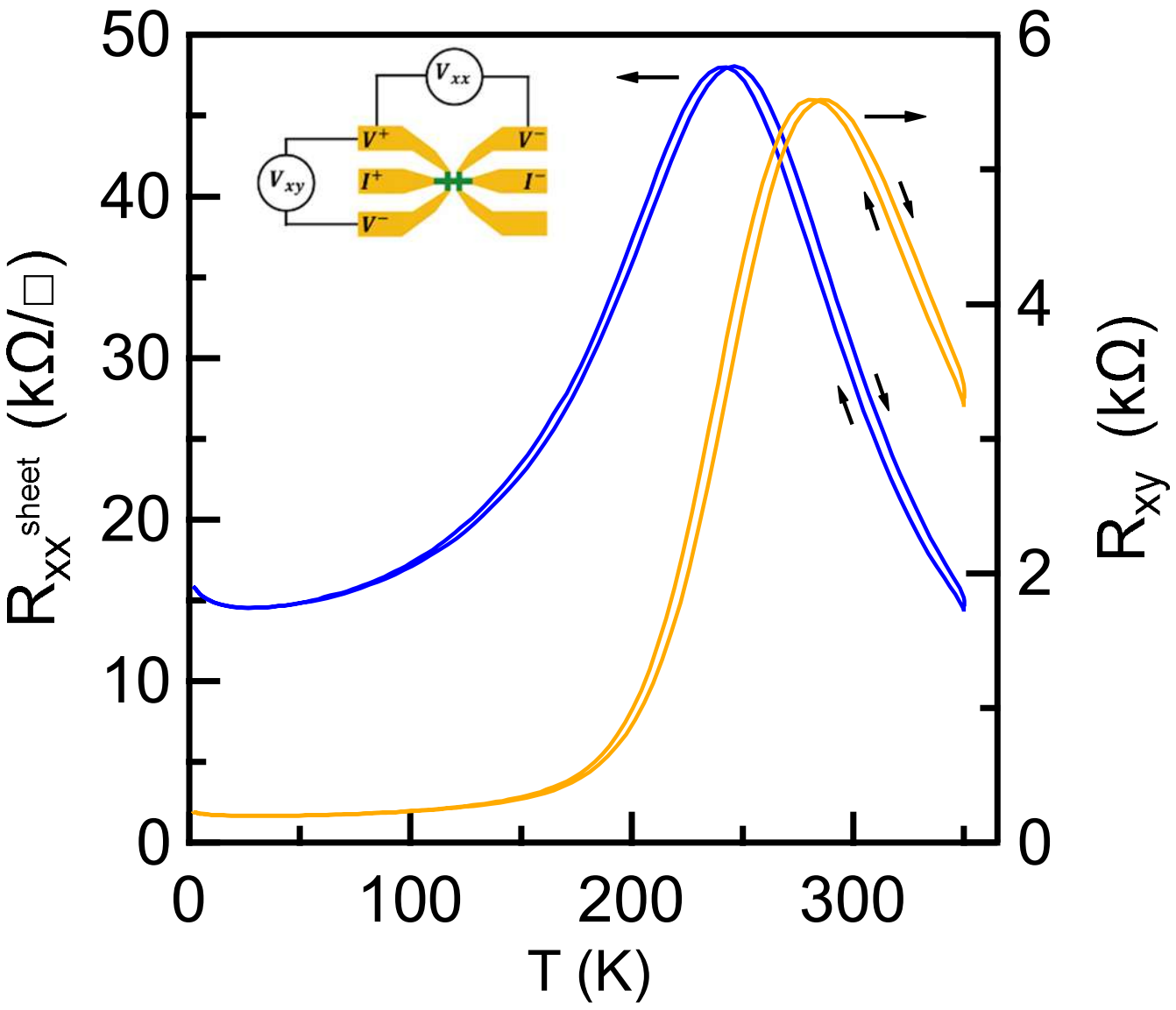

Figure 2. Temperature dependence of longitudinal sheet resistance and Hall resistance of MnTe thin film grown on InP (111). Below 250 K, $R_{xx}$ exhibits a metallic-like behavior ($dR_{xx}/dT > 0$) despite MnTe being a semiconductor. Both channels exhibit peaks at similar but non-overlapping temperatures.

Figure 3 exhibits the anisotropic magnetoresistance (AMR) of MnTe thin films, measured as a function of temperature and magnetic field. AMR could be used as a probe into magnetic structure and its evolution with applied magnetic field in itinerant anti-ferromagnets[35–38]. Figure 3 shows the $AMR(\%) = (R_{xx} - R_{min})/R_{min} \times 100$, where the current is kept along a crystallographic direction as the angle between the current and the in-plane magnetic field changes. AMR response in an itinerant (anti)ferromagnet contains both crystalline and non-crystalline contributions[39,40].

The crystalline component in an itinerant antiferromagnet depends on the angle between the Néel vector and a specific crystallographic direction. The non-crystalline AMR, to first order, depends on how the scattering matrix elements of the charge carriers change with the angle between the spin axis and the momentum[35,39]. While the crystalline component of AMR reflects the underlying lattice symmetry, the non-crystalline AMR component has a two-fold symmetry. Here, the room temperature AMR exhibits a two-fold symmetry. Cooling below the transition temperature, a six-fold AMR symmetry emerges, which reflects the in-plane symmetry of the NiAs-type structure and MnTe/InP (111) heterostructure. We conclude that crystalline AMR is the dominant contribution to observed AMR signal[40–43].

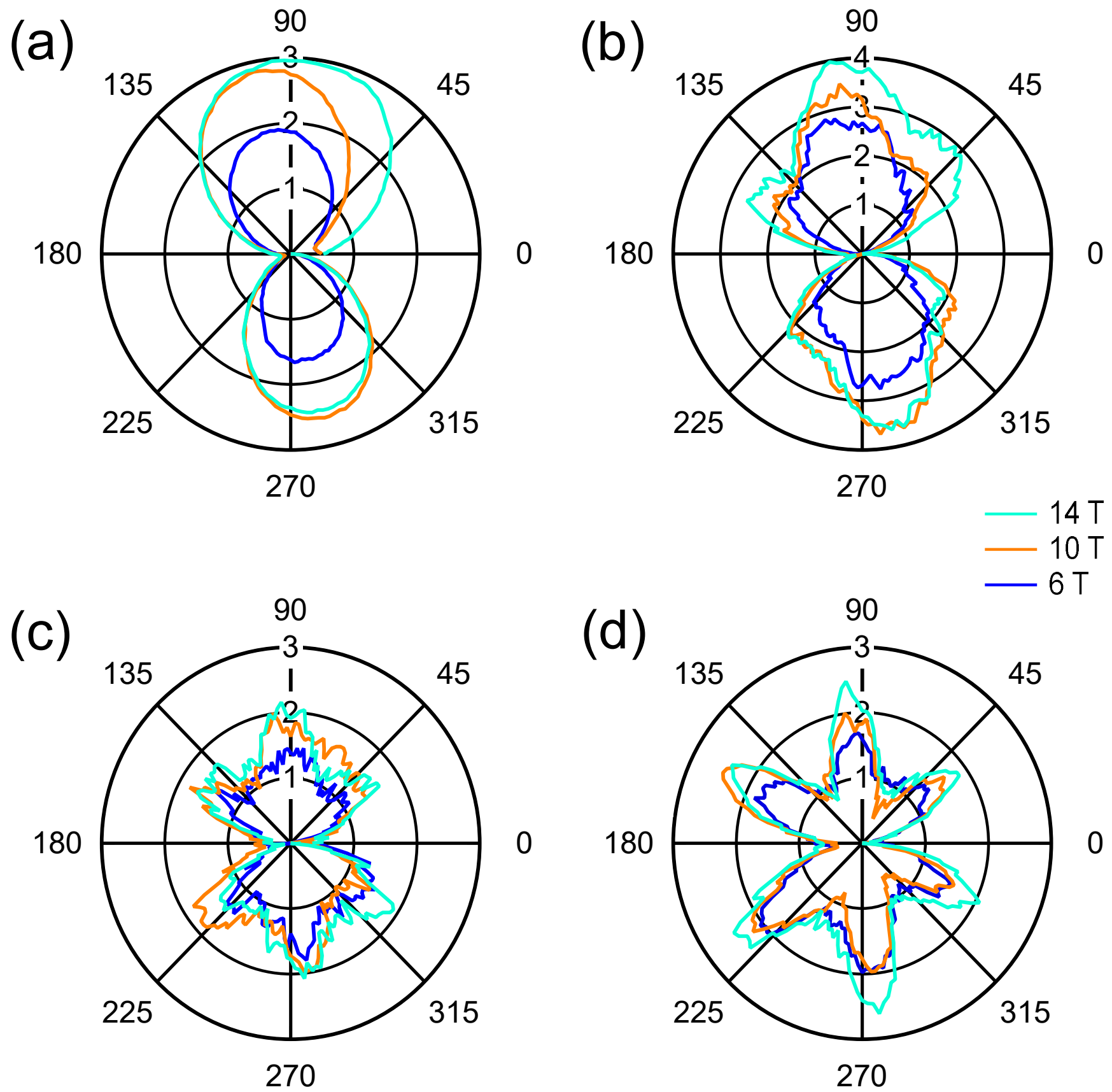

Figure 3. Polar plots showing the angular dependence of anisotropic magnetoresistance ($AMR(\%) = (R_{xx}-R_{min})/R_{min} \times 100$) in MnTe thin films at different temperatures and magnetic fields: (a) 300 K, (b) 150 K, (c) 50 K, and (d) 2 K, under magnetic fields of 6 T, 10 T, and 14 T. At room temperature, the AMR exhibits predominantly two-fold symmetry, whereas upon cooling a six-fold symmetry emerges, reflecting the underlying hexagonal crystal symmetry of the NiAs-type MnTe structure.

Figure 4 (a) and (b) exhibit longitudinal and transverse resistance of MnTe film, respectively, as a function of sweeping magnetic field at 300 K, 200 K, 100 K, 50 K, and 1.8 K. At room temperature, the longitudinal magneto-resistance of MnTe films exhibits a positive parabolic field dependence which is commonly observed in paramagnetic metals and semiconductors[35]. At 200 K a negative magneto-resistance emerges at low fields (<6 T). Furthermore, two resistance minima appear at approximately +6 T and – 6 T, resulting in a W-shape profile. Further reduction of the temperature below 200 K gives rise to butterfly-shaped hysteresis, suggesting magnetic polarization and breaking of time-reversal symmetry [21,40,43]. The hysteresis widens with further reduction of temperature as the magnetic order parameter strengthens. We observe saturation fields reaching 20 T in the longitudinal magneto-resistance measured at 400 mK [Figure S4].

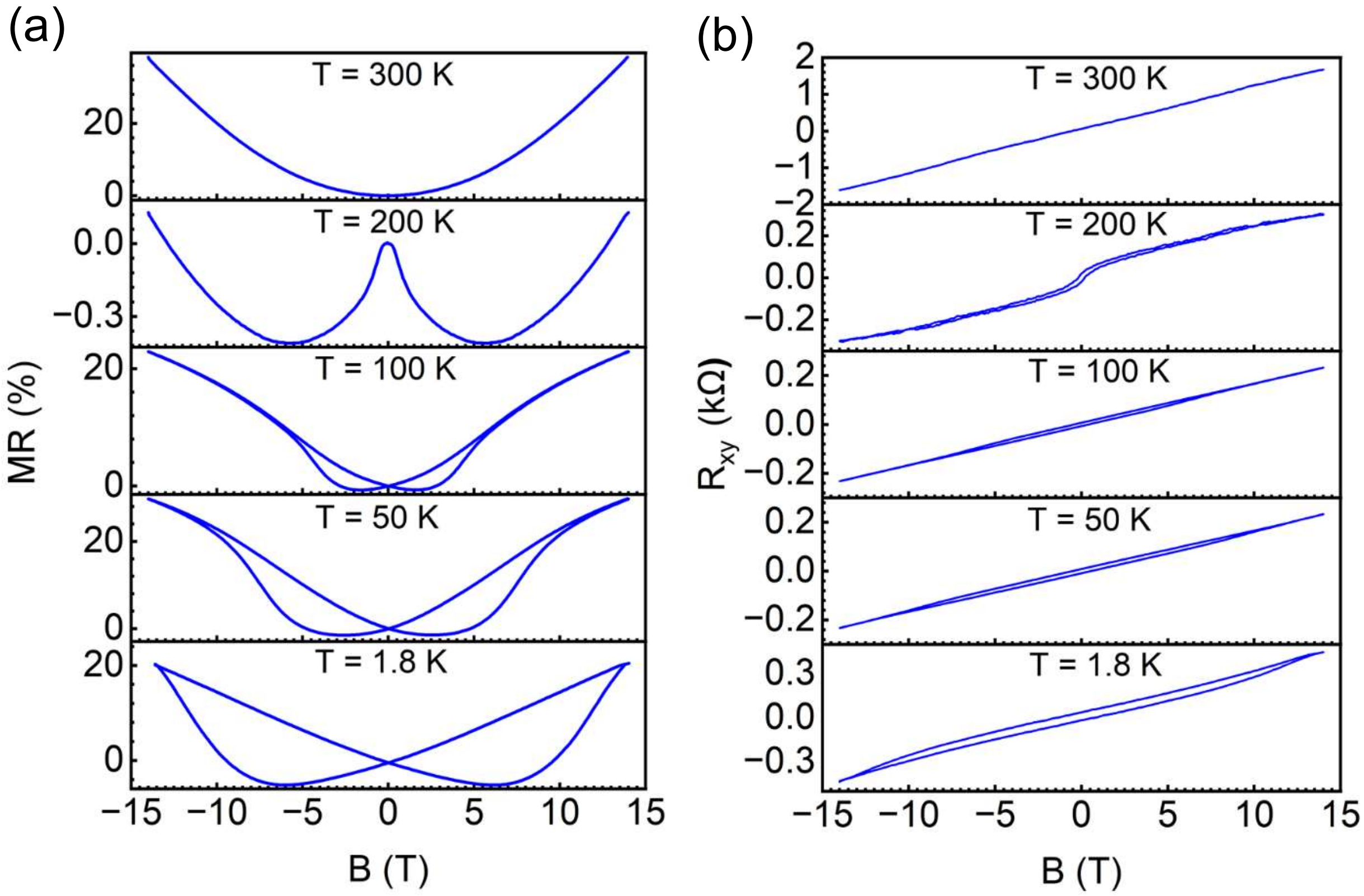

Figure 4. Magnetic field dependence ($MR(\%) = (R - \mathrm{R}_{B=0})/\mathrm{R}_{B=0} \times 100$) of (a) longitudinal, and (b) Transverse magnetoresistance at temperatures of 300, 200, 100, 50, and 1.8 K. With decreasing temperature, both the longitudinal and transverse responses exhibit hysteretic behavior.

The transverse magneto-resistance at room temperature exhibits a linear response with positive slope, suggesting presence of hole-type charge carriers. The behavior becomes hysteretic and non-linear at lower temperatures, consistent with longitudinal magneto-resistance. The emergence of hysteretic butterfly behavior in longitudinal channel and hysteretic non-linear behavior in transverse magneto-resistance, suggest the emergence of a magnetic order parameter which breaks time-reversal symmetry. This behavior is not expected in conventional collinear antiferromagnets, in which symmetry-related antiparallel sublattices enforce spin-degenerate electronic states and could be potentially explained by altermagnetism where the spins remain antiferromagnetically ordered in real space but have a momentum dependent time-reversal symmetry breaking[6,44]. The alternative extrinsic explanation could be presence of defects such as planar defects (e.g., stacking faults and hetero-interfaces) or point defects (e.g., anti-sites and interstitial manganese)[28,45,46]. The magneto-resistance response does not show sensitivity to in-plane transport direction, pointing to extrinsic explanations of the observed magnetic response.

To understand the origin of the magnetic characteristics observed in MnTe thin films, we performed a series of first-principles calculations considering several possible scenarios that may arise during the growth. We first investigated the MnTe/InP(111) heterostructure. The initial MnTe/InP(111) heterostructure considered is shown in Figure S5(a). The MnTe slab consists of four MnTe units, each containing Mn atoms with opposite spin orientations. Although the MnTe layers are magnetically compensated and the InP substrate is nonmagnetic, the combined heterostructure exhibits a finite ferromagnetic moment after structural relaxation and self-consistent calculations. The corresponding band structure and density of states (DOS) are shown in Figures S5(b) and S5(c), respectively. These results demonstrate that magnetic polarization can emerge in MnTe when it is interfaced with another materials. The local environment modifies the magnetic moments and magnetization-density distributions, resulting in incomplete compensation and a finite net magnetization.

We have also explored several defect configurations that may occur during the growth, including Mn substitution at Te sites, Mn interstitials, and Te vacancies [Figure 5]. We employed magnetically compensated initial configurations with equal numbers of spin-up and spin-down Mn atoms to avoid any artificial net magnetization. Our calculations reveal that Mn-related defects, namely Mn substitution and Mn interstitials, promote the emergence of a ferromagnetic state, whereas Te vacancies retain the antiferromagnetic ordering of MnTe. Among these mechanisms, we believe that Mn interstitial defects may be the likely source of observed magnetic polarization. Maintaining an exact 1:1 Mn-to-Te stoichiometric ratio during the growth is experimentally challenging due to volatility of tellurium at growth conditions, and a small amount of excess manganese could be incorporated into the MnTe lattice. These additional Mn atoms may occupy interstitial sites, disturb the magnetic ordering, and generate a finite net magnetization[29,47].

In summary, epitaxial α-MnTe thin films are grown on InP (111) by molecular beam epitaxy. Magneto-transport measurements reveal a hysteretic butterfly longitudinal and nonlinear transverse magneto-resistance responses at low temperatures, indicating a finite net magnetic polarization. Our DFT results suggest that although bulk MnTe is an antiferromagnet, there are multiple mechanisms through which finite magnetization could emerge in MnTe thin films such as interface-induced symmetry breaking and point defects. Among these mechanisms, manganese interstitials may play the dominant role since excess manganese can be incorporated in the MnTe structure.

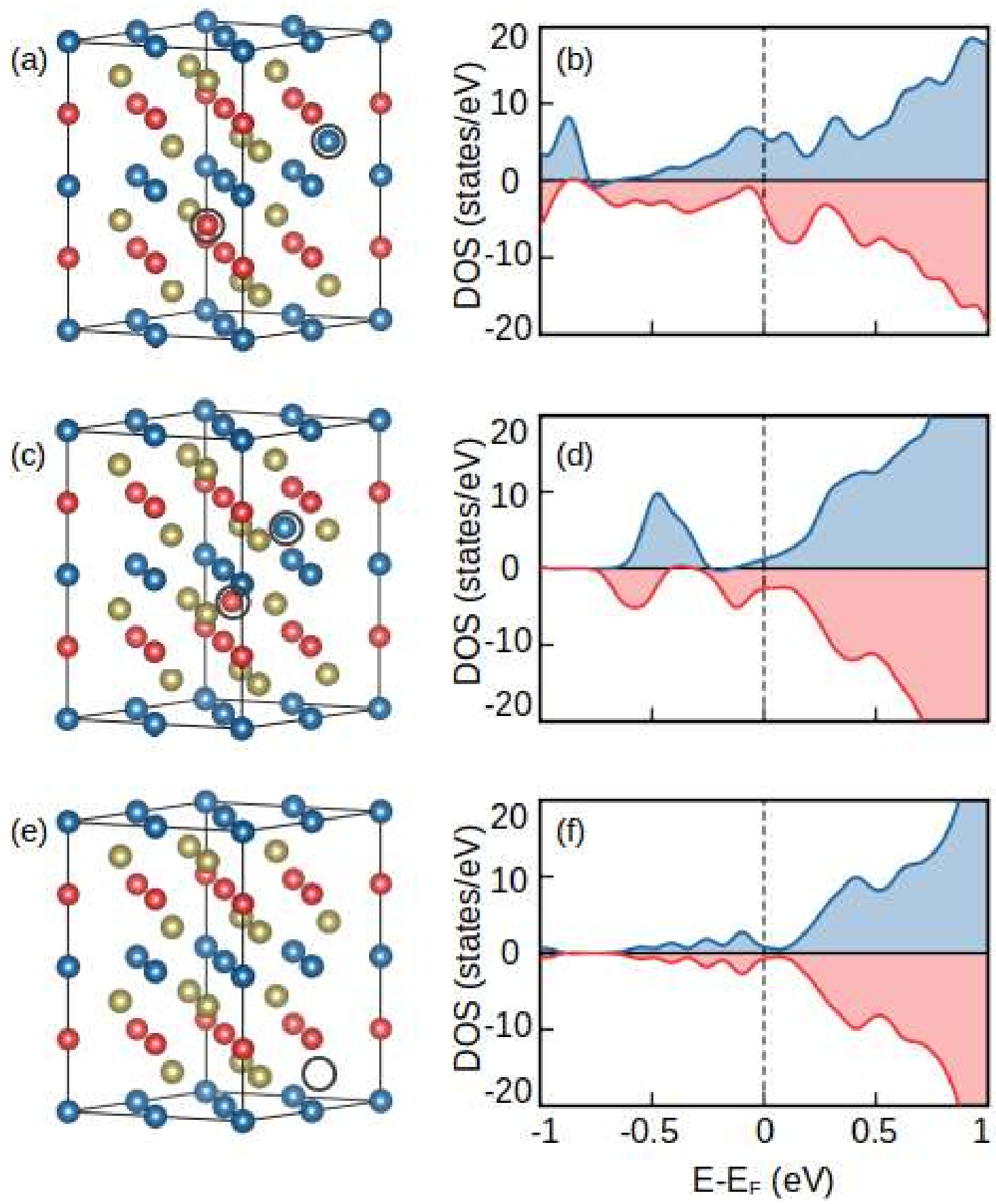

Figure 5. Effect of defects on the magnetic state of MnTe. (a), (c), and (e) show the crystal structures containing two Mn atoms substituting Te sites, two Mn interstitial atoms, and one Te vacancy, respectively. Mn atoms with opposite spin orientations are shown in blue and red spheres, and Te atoms are shown in golden color. The defect locations are highlighted by black circles. The corresponding spin-polarized DOS are shown in (b), (d), and (f). While all three defect configurations preserve overall total magnetic compensation, the substitution of Te by Mn atoms and the introduction of two Mn interstitials induce a pronounced spin asymmetry in the electronic structure, indicating the emergence of a ferromagnetic state. In contrast, the Te-vacancy configuration retains an antiferromagnetic character.

## Acknowledgements

Y.G and S.J.P were supported by the U.S. National Science Foundation under Grant No. NSF DMR-2408890. Partial funding for the OSU NanoSystems Laboratory shared facilities used in this research was provided by the Center for Emergent Materials: an NSF MRSEC under award number DMR-2011876. This study is also supported by the research conducted at The Pennsylvania State University Two-Dimensional Crystal Consortium – Materials Innovation Platform (2DCC-MIP) which is supported by NSF cooperative agreement DMR-2039351. A.N. acknowledges support from Department of Science and Technology Core Research Grant (CRG/2023/000114). A portion of this work was performed at the National High Magnetic Field Laboratory, which is supported by National Science Foundation Cooperative Agreement No. DMR-2128556* and the State of Florida.

## Data Availability Statement

The data that support the findings of this study are available in the article and its Supplemental Material. Raw data are available at https://list.2dccmip.org/list/data/10B-Vrr16X2LP0Xi

## Conflict of Interest Statement

The authors have no conflicts of interest to declare.

## References

(1) Wolf, S. A.; Awschalom, D. D.; Buhrman, R. A.; Daughton, J. M.; von Molnár, S.; Roukes, M. L.; Chtchelkanova, A. Y.; Treger, D. M. Spintronics: A Spin-Based Electronics Vision for the Future. *Science* **2001**, *294* (5546), 1488–1495. https://doi.org/10.1126/science.1065389.

(2) Žutić, I.; Fabian, J.; Das Sarma, S. Spintronics: Fundamentals and Applications. *Rev. Mod. Phys.* **2004**, *76* (2), 323–410. https://doi.org/10.1103/RevModPhys.76.323.

(3) Fert, A.; Cros, V.; Sampaio, J. Skyrmions on the Track. *Nature Nanotech* **2013**, *8* (3), 152–156. https://doi.org/10.1038/nnano.2013.29.

(4) Lee, J. H.; Fang, L.; Vlahos, E.; Ke, X.; Jung, Y. W.; Kourkoutis, L. F.; Kim, J.-W.; Ryan, P. J.; Heeg, T.; Roeckerath, M.; Goian, V.; Bernhagen, M.; Uecker, R.; Hammel, P. C.; Rabe, K. M.; Kamba, S.; Schubert, J.; Freeland, J. W.; Muller, D. A.; Fennie, C. J.; Schiffer, P.; Gopalan, V.; Johnston-Halperin, E.; Schlom, D. G. A Strong Ferroelectric Ferromagnet Created by Means of Spin–Lattice Coupling. *Nature* **2010**, *466* (7309), 954–958. https://doi.org/10.1038/nature09331.

(5) *Unit-cell dimensions of α-MnTe in the 295 K – 1200 K temperature range - Minikayev - 2015 - X-Ray Spectrometry - Wiley Online Library*. https://analyticalsciencejournals.onlinelibrary.wiley.com/doi/10.1002/xrs.2646 (accessed 2026-07-10).

(6) Devaraj, N. Interplay of Altermagnetism and Pressure in Hexagonal and Orthorhombic MnTe. *Phys. Rev. Mater.* **2024**, *8* (10), 104407. https://doi.org/10.1103/PhysRevMaterials.8.104407.

(7) Yang, L.; Wang, Z. H.; Zhang, Z. D. Electrical Properties of NiAs-Type MnTe Films with Preferred Crystallographic Plane of (110). *J. Appl. Phys.* **2016**, *119* (4), 045304. https://doi.org/10.1063/1.4940950.

(8) Krause, M.; Bechstedt, F. Structural and Magnetic Properties of MnTe Phases from Ab Initio Calculations. *J Supercond Nov Magn* **2013**, *26* (5), 1963–1972. https://doi.org/10.1007/s10948-012-2071-6.

(9) Mazin, I. I. Altermagnetism in MnTe: Origin, Predicted Manifestations, and Routes to Detwinning. *Phys. Rev. B* **2023**, *107* (10), L100418. https://doi.org/10.1103/PhysRevB.107.L100418.

(10) Šmejkal, L.; Sinova, J.; Jungwirth, T. Emerging Research Landscape of Altermagnetism. *Phys. Rev. X* **2022**, *12* (4), 040501. https://doi.org/10.1103/PhysRevX.12.040501.

(11) Osumi, T.; Souma, S.; Aoyama, T.; Yamauchi, K.; Honma, A.; Nakayama, K.; Takahashi, T.; Ohgushi, K.; Sato, T. Observation of a Giant Band Splitting in Altermagnetic MnTe. *Phys. Rev. B* **2024**, *109* (11), 115102. https://doi.org/10.1103/PhysRevB.109.115102.

(12) Hajlaoui, M.; Wilfred D'Souza, S.; Šmejkal, L.; Kriegner, D.; Krizman, G.; Zakusylo, T.; Olszowska, N.; Caha, O.; Michalička, J.; Sánchez-Barriga, J.; Marmodoro, A.; Výborný, K.; Ernst, A.; Cinchetti, M.; Minar, J.; Jungwirth, T.; Springholz, G. Temperature Dependence of Relativistic Valence Band Splitting Induced by an Altermagnetic Phase Transition. *Advanced Materials* **2024**, *36* (31), 2314076. https://doi.org/10.1002/adma.202314076.

(13) Zhang, Q.; Jensen, C. J.; Grutter, A. J.; Santhosh, S.; Ratcliff, W. D.; Borchers, J. A.; Heitmann, T. W.; Narayanan, N.; Charlton, T. R.; Yu, M.; Wang, K.; Auker, W.; Samarth, N.; Law, S. Examining the Spin Structure of Altermagnetic Candidate MnTe Grown with Near Ideal Stoichiometry. arXiv October 24, 2025. https://doi.org/10.48550/arXiv.2510.21511.

(14) Giannozzi, P.; Baroni, S.; Bonini, N.; Calandra, M.; Car, R.; Cavazzoni, C.; Ceresoli, D.; Chiarotti, G. L.; Cococcioni, M.; Dabo, I.; Dal Corso, A.; de Gironcoli, S.; Fabris, S.; Fratesi, G.; Gebauer, R.; Gerstmann, U.; Gougoussis, C.; Kokalj, A.; Lazzeri, M.; Martin-Samos, L.; Marzari, N.; Mauri, F.; Mazzarello, R.; Paolini, S.; Pasquarello, A.; Paulatto, L.; Sbraccia, C.; Scandolo, S.; Sclauzero, G.; Seitsonen, A. P.; Smogunov, A.; Umari, P.; Wentzcovitch, R. M. QUANTUM ESPRESSO: A Modular and Open-Source Software Project for Quantum Simulations of Materials. *J. Phys.: Condens. Matter* **2009**, *21* (39), 395502. https://doi.org/10.1088/0953-8984/21/39/395502.

(15) Blöchl, P. E. Projector Augmented-Wave Method. *Phys. Rev. B* **1994**, *50* (24), 17953–17979. https://doi.org/10.1103/PhysRevB.50.17953.

(16) Perdew, J. P.; Burke, K.; Ernzerhof, M. Generalized Gradient Approximation Made Simple. *Phys. Rev. Lett.* **1996**, *77* (18), 3865–3868. https://doi.org/10.1103/PhysRevLett.77.3865.

(17) Cococcioni, M.; de Gironcoli, S. Linear Response Approach to the Calculation of the Effective Interaction Parameters in the LDA+U Method. *Phys. Rev. B* **2005**, *71* (3), 035105. https://doi.org/10.1103/PhysRevB.71.035105.

(18) Gonzalez Betancourt, R. D.; Zubáč, J.; Gonzalez-Hernandez, R.; Geishendorf, K.; Šobáň, Z.; Springholz, G.; Olejník, K.; Šmejkal, L.; Sinova, J.; Jungwirth, T.; Goennenwein, S. T. B.; Thomas, A.; Reichlová, H.; Železný, J.; Kriegner, D. Spontaneous Anomalous Hall Effect Arising from an Unconventional Compensated Magnetic Phase in a Semiconductor. *Phys. Rev. Lett.* **2023**, *130* (3), 036702. https://doi.org/10.1103/PhysRevLett.130.036702.

(19) Rooj, S.; Chakraborty, J.; Ganguli, N. Hexagonal MnTe with Antiferromagnetic Spin Splitting and Hidden Rashba–Dresselhaus Interaction for Antiferromagnetic Spintronics. *Advanced Physics Research* **2024**, *3* (1), 2300050. https://doi.org/10.1002/apxr.202300050.

(20) Szuszkiewicz, W.; Dynowska, E.; Witkowska, B.; Hennion, B. Spin-Wave Measurements on Hexagonal MnTe of NiAs-Type Structure by Inelastic Neutron Scattering. *Phys. Rev. B* **2006**, *73* (10), 104403. https://doi.org/10.1103/PhysRevB.73.104403.

(21) Devaraj, N.; Bose, A.; Das, A.; Reja, M. A.; Mandal, A.; Narayan, A.; Nanda, B. R. K. Unlocking Doping Effects on Altermagnetism in MnTe: Emergence of Quasi-Altermagnetism. *Phys. Rev. B* **2026**, *113* (10), 104438. https://doi.org/10.1103/k36v-91br.
(22) Watanabe, R.; Yoshimi, R.; Shirai, M.; Tanigaki, T.; Kawamura, M.; Tsukazaki, A.; Takahashi, K. S.; Arita, R.; Kawasaki, M.; Tokura, Y. Emergence of Interfacial Conduction and Ferromagnetism in MnTe/InP. *Appl. Phys. Lett.* **2018**, *113* (18), 181602. https://doi.org/10.1063/1.5050446.
(23) Siol, S.; Han, Y.; Mangum, J.; Schulz, P.; Holder, A. M.; Klein, T. R.; van Hest, M. F. A. M.; Gorman, B.; Zakutayev, A. Stabilization of Wide Band-Gap p-Type Wurtzite MnTe Thin Films on Amorphous Substrates. *J. Mater. Chem. C* **2018**, *6* (23), 6297–6304. https://doi.org/10.1039/c8tc01828f.
(24) Cao, H.; Zhang, J.; Bai, W.; Zhao, D.; Lin, R.; Wang, X.; Yang, J.; Zhang, Y.; Qi, R.; Huang, R.; Tang, X.; Wang, J.; Chu, J. Molecular Beam Epitaxy, Photoluminescence from Mn2+ Multiplets and Self-Trapped Exciton States of γ-MnTe Single-Crystalline Thin Films. *Applied Surface Science* **2023**, *611*, 155733. https://doi.org/10.1016/j.apsusc.2022.155733.
(25) Mori, S.; Sutou, Y.; Ando, D.; Koike, J. Optical and Electrical Properties of α-MnTe Thin Films Deposited Using RF Magnetron Sputtering. *Materials Transactions* **2018**, *59* (9), 1506–1512. https://doi.org/10.2320/matertrans.M2018086.
(26) Arnault, E. G.; Al-Tawhid, A. H.; Salmani-Rezaie, S.; Muller, D. A.; Kumah, D. P.; Bahramy, M. S.; Finkelstein, G.; Ahadi, K. Anisotropic Superconductivity at KTaO3(111) Interfaces. *Science Advances* **2023**, *9* (7), eadf1414. https://doi.org/10.1126/sciadv.adf1414.
(27) Al-Tawhid, A. H.; Poage, S. J.; Salmani-Rezaie, S.; Gonzalez, A.; Chikara, S.; Muller, D. A.; Kumah, D. P.; Gastiasoro, M. N.; Lorenzana, J.; Ahadi, K. Enhanced Critical Field of Superconductivity at an Oxide Interface. *Nano Lett.* **2023**, *23* (15), 6944–6950. https://doi.org/10.1021/acs.nanolett.3c01571.
(28) Bey, S.; Zhukovskyi, M.; Orlova, T.; Fields, S.; Lauter, V.; Ambaye, H.; Ievlev, A.; Bennett, S. P.; Liu, X.; Assaf, B. A. Interface, Bulk and Surface Structure of Heteroepitaxial Altermagnetic α-MnTe Films Grown on GaAs(111). *Phys. Rev. Mater.* **2025**, *9* (7), 074404. https://doi.org/10.1103/1dh1-mly1.
(29) Chilcote, M.; Mazza, A. R.; Lu, Q.; Gray, I.; Tian, Q.; Deng, Q.; Moseley, D.; Chen, A.-H.; Lapano, J.; Gardner, J. S.; Eres, G.; Ward, T. Z.; Feng, E.; Cao, H.; Lauter, V.; McGuire, M. A.; Hermann, R.; Parker, D.; Han, M.-G.; Kayani, A.; Rimal, G.; Wu, L.; Charlton, T. R.; Moore, R. G.; Brahlek, M. Stoichiometry-Induced Ferromagnetism in Altermagnetic Candidate MnTe. *Advanced Functional Materials* **2024**, *34* (46), 2405829. https://doi.org/10.1002/adfm.202405829.
(30) Efrem D'Sa, J. B. C.; Bhobe, P. A.; Priolkar, K. R.; Das, A.; Paranjpe, S. K.; Prabhu, R. B.; Sarode, P. R. Low-Temperature Neutron Diffraction Study of MnTe. *Journal of Magnetism and Magnetic Materials* **2005**, *285* (1), 267–271. https://doi.org/10.1016/j.jmmm.2004.08.001.
(31) Pereverzev, Yu. V.; Povstyanyi, L. V.; Zvyagin, A. I. Electrical and Optical Properties of the Magnetic Semiconductor MnTe. *Sov. J. Low Temp. Phys.* **1976**, *2* (5), 311–313. https://doi.org/10.1063/10.0029306.
(32) Aoyama, T.; Ohgushi, K. Piezomagnetic Properties in Altermagnetic MnTe. *Phys. Rev. Mater.* **2024**, *8* (4), L041402. https://doi.org/10.1103/PhysRevMaterials.8.L041402.
(33) Kl/osowski, P.; Giebul/towicz, T. M.; Rhyne, J. J.; Samarth, N.; Luo, H.; Furdyna, J. K. Antiferromagnetism in Epilayers and Superlattices Containing Zinc-blende MnSe and MnTe. *J. Appl. Phys.* **1991**, *70* (10), 6221–6223. https://doi.org/10.1063/1.350001.
(34) Prechtl, G.; Heiss, W.; Bonanni, A.; Sitter, H.; Jantsch, W.; Mackowski, S.; Karczewski, G. Antiferromagnetic Phase Transition in a Single MnTe Monolayer. *Physica E: Low-dimensional*

*Systems and Nanostructures* **2000**, *7* (3), 1006–1010. https://doi.org/10.1016/S1386-9477(00)00105-3.
(35) Marti, X.; Fina, I.; Frontera, C.; Liu, J.; Wadley, P.; He, Q.; Paull, R. J.; Clarkson, J. D.; Kudrnovský, J.; Turek, I.; Kuneš, J.; Yi, D.; Chu, J.-H.; Nelson, C. T.; You, L.; Arenholz, E.; Salahuddin, S.; Fontcuberta, J.; Jungwirth, T.; Ramesh, R. Room-Temperature Antiferromagnetic Memory Resistor. *Nature Mater* **2014**, *13* (4), 367–374. https://doi.org/10.1038/nmat3861.
(36) Ahadi, K.; Lu, X.; Salmani-Rezaie, S.; Marshall, P. B.; Rondinelli, J. M.; Stemmer, S. Anisotropic Magnetoresistance in the Itinerant Antiferromagnetic EuTiO3. *Phys. Rev. B* **2019**, *99* (4), 041106. https://doi.org/10.1103/PhysRevB.99.041106.
(37) Wang, C.; Seinige, H.; G. Cao; Zhou, J.-S.; Goodenough, J. B.; Tsoi, M. Anisotropic Magnetoresistance in Antiferromagnetic Sr2IrO4. *Phys. Rev. X* **2014**, *4* (4), 041034. https://doi.org/10.1103/PhysRevX.4.041034.
(38) Lu, C.; Gao, B.; Wang, H.; Wang, W.; Yuan, S.; Dong, S.; Liu, J.-M. Revealing Controllable Anisotropic Magnetoresistance in Spin–Orbit Coupled Antiferromagnet Sr2IrO4. *Advanced Functional Materials* **2018**, *28* (17), 1706589. https://doi.org/10.1002/adfm.201706589.
(39) Kriegner, D.; Výborný, K.; Olejník, K.; Reichlová, H.; Novák, V.; Marti, X.; Gazquez, J.; Saidl, V.; Němec, P.; Volobuev, V. V.; Springholz, G.; Holý, V.; Jungwirth, T. Multiple-Stable Anisotropic Magnetoresistance Memory in Antiferromagnetic MnTe. *Nat Commun* **2016**, *7* (1), 11623. https://doi.org/10.1038/ncomms11623.
(40) Gonzalez Betancourt, R. D.; Zubáč, J.; Geishendorf, K.; Ritzinger, P.; Růžičková, B.; Kotte, T.; Železný, J.; Olejník, K.; Springholz, G.; Büchner, B.; Thomas, A.; Výborný, K.; Jungwirth, T.; Reichlová, H.; Kriegner, D. Anisotropic Magnetoresistance in Altermagnetic MnTe. *npj Spintronics* **2024**, *2* (1), 45. https://doi.org/10.1038/s44306-024-00046-z.
(41) Kriegner, D.; Reichlova, H.; Grenzer, J.; Schmidt, W.; Ressouche, E.; Godinho, J.; Wagner, T.; Martin, S. Y.; Shick, A. B.; Volobuev, V. V.; Springholz, G.; Holý, V.; Wunderlich, J.; Jungwirth, T.; Výborný, K. Magnetic Anisotropy in Antiferromagnetic Hexagonal MnTe. *Phys. Rev. B* **2017**, *96* (21), 214418. https://doi.org/10.1103/PhysRevB.96.214418.
(42) Ritzinger, P.; Výborný, K. Anisotropic Magnetoresistance: Materials, Models and Applications. *R Soc Open Sci* **2023**, *10* (10), 230564. https://doi.org/10.1098/rsos.230564.
(43) Orlova, N. N.; Esin, V. D.; Timonina, A. V.; Kolesnikov, N. N.; Deviatov, E. V. Magnetization Symmetry for the Altermagnetic Candidate MnTe. *Phys. Rev. B* **2025**, *111* (22), 224414. https://doi.org/10.1103/br1r-bjzk.
(44) Liu, Z.; Xu, S.; DeStefano, J. M.; Rosenberg, E.; Zhang, T.; Li, J.; Stone, M. B.; Ye, F.; Tian, W.; Edwards, S.; Cong, R.; Pan, S.; Chu, C.-W.; Deng, L.; Morosan, E.; Fernandes, R. M.; Chu, J.-H.; Dai, P. Strain-Tunable Anomalous Hall Effect in Hexagonal MnTe. arXiv March 4, 2026. https://doi.org/10.48550/arXiv.2509.19582.
(45) Kim, W.; Park, I. J.; Kim, H. J.; Lee, W.; Kim, S. J.; Kim, C. S. Room-Temperature Ferromagnetic Property in MnTe Semiconductor Thin Film Grown by Molecular Beam Epitaxy. *IEEE Transactions on Magnetics* **2009**, *45* (6), 2424–2427. https://doi.org/10.1109/TMAG.2009.2018596.
(46) Brand, E.; Rosendal, V.; Wu, Y.; Tran, T.; Palliotto, A.; Maznichenko, I. V.; Ostanin, S.; Esposito, V.; Ernst, A.; Zhou, S.; Park, D.-S.; Pryds, N. Defect-Induced Magnetic Symmetry Breaking in Oxide Materials. *Appl. Phys. Rev.* **2025**, *12* (1), 011327. https://doi.org/10.1063/5.0216796.
(47) Negi, P.; Aradhyula, S. S. S. A.; Dutta, S.; Roychowdhury, S. MnTe Rising: A Modern Era of Thermoelectric and Altermagnetic Materials. *Chem. Mater.* **2025**, *37* (16), 6097–6117. https://doi.org/10.1021/acs.chemmater.5c01561.